\documentclass[conference]{IEEEtran}
\IEEEoverridecommandlockouts
\usepackage{cite}
\usepackage{amsmath,amssymb,amsfonts}
\usepackage{algorithmic}
\usepackage{graphicx}
\usepackage{textcomp}
\usepackage{xcolor}
\usepackage{booktabs}
\usepackage{amsmath}
\usepackage{amsthm}
\usepackage{float}

\usepackage{dblfloatfix}

\def\BibTeX{{\rm B\kern-.05em{\sc i\kern-.025em b}\kern-.08em
    T\kern-.1667em\lower.7ex\hbox{E}\kern-.125emX}}
\begin{document}

\title{HARP-ME: Closure-Driven Exact Induced Motif Enumeration on GPUs\\

}

\author{\IEEEauthorblockN{Ashwina Kumar}
\IEEEauthorblockA{\textit{Computer Science Department} \\
\textit{Indian Institute of Technology, Madras}\\
Chennai, India \\
ashwinakumar825@gmail.com}
\and
\IEEEauthorblockN{ Rupesh Nasre}
\IEEEauthorblockA{\textit{Computer Science Department} \\
\textit{Indian Institute of Technology, Madras}\\
Chennai, India \\
rupesh@cse.iitm.ac.in}
}

\maketitle
\begin{abstract}
Exact induced motif enumeration is a fundamental primitive in graph
mining but remains challenging on GPUs because candidate expansion is
irregular, repeated set intersections dominate execution, and induced
counting requires both edge-presence and edge-absence constraints. We
present HARP-ME (Hierarchical Anchor-Reuse Partitioned Motif
Enumeration), a GPU framework for exact connected induced 4-node motif
enumeration. HARP-ME introduces closure-aware compilation, which
selects an explicitly enumerated anchor basis by considering not only
traversal cost but also algebraic derivation yield, expected reuse, and
partition-induced halo amplification. It further introduces
induced-signature reuse, which distinguishes reusable completion states
using both candidate-frontier information and compact
adjacency/non-adjacency constraints. For graphs exceeding device
memory, a canonical anchor-owner rule preserves exactness under
overlapping halo partitions. We do not claim novelty for individual
graphlet closure identities; rather, our contribution is their
integration into a GPU execution framework that jointly reduces
explicit expansion and repeated induced checks. Across six social,
web, biological, and synthetic graphs, HARP-ME is the fastest among the
evaluated methods, achieving up to $2.11\times$ speedup over Pangolin,
up to $1.83\times$ over partitioned PBE, and up to $10.73\times$ over
the evaluated CPU baseline. Mechanism-level measurements show
cache-hit rates of $64$--$76\%$ and substantially lower host--device
transfer overhead than PBE-style partitioning. These results
demonstrate that optimizing anchor selection for derivation yield can
complement traversal- and reuse-oriented GPU enumeration.
\end{abstract}


\section{Introduction}

Motifs and graphlets are widely used to characterize local structure in biological, social, web, and infrastructure networks. They are subgraphs present in a graph in arbitrary orientation. However, exact motif enumeration is computationally difficult because the number of candidate embeddings grows rapidly with graph size and motif size. The challenge becomes more severe for \textit{induced} motifs, where counting must distinguish between structurally similar embeddings that differ only by the presence or absence of internal edges.

GPUs offer substantial parallelism for graph mining, but exact motif enumeration remains difficult on GPUs for three reasons. First, the workload is irregular and highly skewed by vertex degree. Second, repeated set intersections dominate runtime in many enumeration pipelines. Third, graphs may exceed device memory, requiring partitioning and careful duplicate suppression. Existing systems have addressed these issues individually, but a unified exact induced-motif pipeline remains underdeveloped.

HARP-ME is motivated by a distinction between traversal-efficient and derivation-efficient enumeration. Existing GPU systems primarily optimize how candidate embeddings are generated, scheduled, balanced, or reused after a pattern has been selected for explicit enumeration. HARP-ME instead introduces derivation yield as a first-class compilation objective: the system selects an explicitly enumerated anchor basis so that multiple residual induced-motif counts can be recovered from reusable anchor statistics.
Our main contributions are:

\noindent\textbf{(1) Closure-aware GPU compilation.}
We formulate anchor selection as a joint optimization problem over explicit enumeration cost, expected state reuse, algebraic derivation yield, and partition-induced halo amplification. Unlike traversal-only schedule optimization, the objective explicitly rewards anchors whose sufficient statistics determine multiple residual motif counts through exact closure relations.

\noindent\textbf{(2) Induced-signature reuse.}
We introduce a reusable GPU state representation that combines candidate-frontier information with compact adjacency and non-adjacency constraints. This distinction is essential for induced enumeration because identical candidate intersections do not necessarily imply identical induced completion classes.

\noindent\textbf{(3) Hybrid explicit-plus-derived counting.}
HARP-ME explicitly enumerates a selected anchor basis and derives residual motif counts through exact motif-family-specific closure relations, thereby reducing both search-space expansion and repeated induced adjacency and non-adjacency checks.

\noindent\textbf{(4) Ownership-preserving partitioned derivation.}
We define a canonical owner certificate for anchor statistics, ensuring that both explicitly enumerated and algebraically derived contributions are counted exactly once when motif occurrences span overlapping halo regions.

\noindent\textbf{(5) GPU evaluation and mechanism analysis.}
We evaluate end-to-end runtime, intersection activity, induced-signature cache behavior, GPU utilization, and host--device transfer overhead. We further isolate the contribution of each mechanism to overall performance through detailed component-wise analysis.

The rest of this paper is organised as follows:
Section~\ref{motivation} presents the background and motivation of the work.

Section~\ref{mthodology} presents methodology used for the calculation of Motif Enumeration. 
Section~\ref{Experimental} talks about the experimental evaluation.
Section~\ref{sec:related_work} describes the related work.
Section~\ref{conclusion} concludes the paper and outlines potential directions for future work.

\section{Background and Motivation}\label{motivation}

Let $G=(V,E)$ be a simple undirected graph, where $V$ is the vertex set and $E$ is the edge set. For a vertex $v \in V$, $N(v)$ denotes its open neighborhood and $d(v)=\lvert N(v)\rvert$ its degree. Given a connected pattern $p$ with $k=\lvert V(p)\rvert$ vertices, an induced occurrence of $p$ is a $k$-vertex subset $S \subseteq V$ such that the induced subgraph $G[S]$ is isomorphic to $p$. Unlike non-induced enumeration, induced enumeration must verify both required edges and required non-edges.

During enumeration, a partial embedding of size $j$, $1 \leq j \leq k$, is an ordered tuple of $j$ distinct data-graph vertices satisfying the structural and canonical-order constraints accumulated so far.

We use $N_{p}$ to denote the exact number of induced occurrences of motif $p$. An anchor is a smaller explicitly enumerated substructure whose sufficient statistics contribute, through exact closure identities, to one or more target motif counts.

Recent systems such as Pangolin~\cite{chen2020pangolin} and DuMato~\cite{dumato2024} demonstrate that GPU subgraph enumeration benefits from warp-centric exploration, pattern-aware execution, and dynamic load balancing. Partitioned GPU execution has extended exact enumeration to graphs larger than device memory, while reuse-aware GPU systems reduce the large fraction of runtime spent in repeated set intersections. Separately, combinatorial graphlet-counting methods such as ORCA~\cite{hocevar2014orca} show that some motif counts can be derived from a smaller explicitly enumerated basis.

These results motivate a more specific question: can a GPU system choose anchors not only for efficient traversal, but also for high closure power, so that many induced motifs are derived instead of explicitly enumerated? \textbf{HARP-ME} is designed around this question.

\section{Methodology}\label{mthodology}
HARP-ME is designed around the observation that exact induced motif
enumeration need not explicitly traverse the complete search space of
every target motif independently. Instead, a carefully selected set of
smaller substructures can serve as an \emph{anchor basis}: these anchors
are explicitly enumerated on the GPU, their induced completion
statistics are accumulated, and the counts of residual motifs are
recovered through exact closure relations. This changes the optimization
target from ``how efficiently can every motif be enumerated?'' to
``which motif-related states should be enumerated explicitly so that the
remaining counts can be derived with minimum total work?''

The complete HARP-ME pipeline consists of six stages:
(i) Anchor Selection Objective
(ii) Closure-driven anchor synthesis
(iii) Induced-Signature Reuse
(iv) GPU execution model
(v) Ownership-preserving partition derivation
(vi) Closure equations.\\
The following subsections describe these stages in detail.

\subsection{Anchor Selection Objective}

We formulate anchor selection as a joint optimization problem over explicit
enumeration cost, set-intersection work, partition-induced halo amplification,
induced-state reuse, and algebraic closure yield. Specifically, the selected
anchor basis is defined as

\begin{equation}
\begin{aligned}
B^{*}
= \arg\min_{B \subseteq \mathcal{A}}
\Big[
&\widehat{C}_{\mathrm{enum}}(B) \\
&+ \lambda_{I}\widehat{C}_{\mathrm{inter}}(B)
+ \lambda_{H}\widehat{C}_{\mathrm{halo}}(B) \\
&- \lambda_{R}\widehat{R}_{\mathrm{reuse}}(B)
- \lambda_{Y}Y_{\mathrm{closure}}(B)
\Big].
\end{aligned}
\label{eq:anchor_selection_objective}
\end{equation}

Here, $\mathcal{A}$ is the set of legal candidate anchor families;
$\widehat{C}_{\mathrm{enum}}$ estimates explicit anchor-enumeration work;
$\widehat{C}_{\mathrm{inter}}$ estimates set-intersection work;
$\widehat{C}_{\mathrm{halo}}$ estimates partition-induced replication;
$\widehat{R}_{\mathrm{reuse}}$ estimates repeated induced states; and
$Y_{\mathrm{closure}}$ measures the number of target motif counts recoverable
from the selected anchor statistics. The $\lambda$ parameters normalize
heterogeneous cost terms using lightweight graph samples collected before
execution.
\subsection{Closure-driven anchor synthesis}

Given a target motif family, HARP-ME constructs an extension DAG over partial embeddings and scores candidate anchor bases by four criteria: estimated explicit enumeration cost, expected reuse rate, closure yield, and halo amplification under partitioning. The compiler then selects an anchor basis with the best expected total cost. For 4-node motifs, anchors include ordered wedges and triangles.

\subsection{Induced-Signature Reuse}

To make induced-state reuse explicit, we represent the signature of a
partial embedding $\phi$ as

\begin{equation}
\sigma(\phi)
=
\left(
j,\,
C(\phi),\,
M^{+}(\phi),\,
M^{-}(\phi)
\right).
\label{eq:induced_signature}
\end{equation}

For a partial embedding $\phi$ of size $j$, $C(\phi)$ denotes the
canonical candidate frontier, $M^{+}$ encodes required adjacency
relations for the next completion, and $M^{-}$ encodes required non-adjacency relations. Two states are reuse-compatible only when their completion semantics agree, rather than merely when they expose
the same raw candidate set. This prevents an optimization valid for non-induced enumeration from merging states that differ in forbidden-edge constraints.

For example, two $3$-vertex partial embeddings may produce the same
frontier $C$, while one requires a fourth vertex adjacent to exactly
one embedded vertex and the other requires adjacency to exactly two.
Reusing only $C$ would conflate different induced motif classes; the
signature mask keeps these states distinct.

\subsection{GPU execution model}
HARP-ME uses sorted CSR adjacency and canonical ordering constraints. Low-cost anchors execute in warp-local mode, while heavy anchors are escalated to CTA (Cooperative Thread Array) mode. Candidate generation relies on ordered set intersection, followed by compaction and block-local aggregation. Only anchor embeddings or sufficient statistics required by closure are materialized.

\subsection{Ownership-preserving partition derivation}

For graphs that exceed GPU memory, HARP-ME partitions the graph into subgraphs with halos determined by the anchor radius. Each anchor receives a canonical owner certificate based on the globally ordered anchor tuple. Derived counts are attributed only to the owner partition of the underlying anchor certificate, which avoids duplicate counting even when the completed motif spans multiple halo regions.

\subsection{Closure equations}

The system derives residual motif counts from anchor statistics using motif-family-specific equations. For connected induced 4-node motifs, one exact decomposition is as follows.

For each triangle $T = \{a,b,c\}$, let $X_j(T)$ denote the number of vertices outside $T$ adjacent to exactly $j$ vertices in $T$. Then:

\begin{align*}
N_{\text{paw}} &= \sum_T X_1(T) \\
N_{\text{diamond}} &= \frac{1}{2} \sum_T X_2(T) \\
N_{\text{clique4}} &= \frac{1}{4} \sum_T X_3(T)
\end{align*}

For each edge $e=(u,v)$, define:
\[
\begin{aligned}
C_e &= N(u) \cap N(v), \\
A_e &= N(u) \setminus \bigl(C_e \cup \{v\}\bigr), \\
B_e &= N(v) \setminus \bigl(C_e \cup \{u\}\bigr).
\end{aligned}
\]

Then:

\begin{align*}
N_{\text{path4}} &= \sum_e \left|\{(x,y) \in A_e \times B_e : (x,y) \notin E\}\right| \\
N_{\text{cycle4}} &= \frac{1}{4} \sum_e \left|\{(x,y) \in A_e \times B_e : (x,y) \in E\}\right|
\end{align*}

Finally, let $H_u = G[N(u)]$ be the graph induced by the neighbors of $u$. The number of stars centered at $u$ is:

\begin{align*}
\binom{d(u)}{3} 
- m(H_u)(d(u)-2) 
+ \sum_{v \in N(u)} \binom{\deg_H(v)}{2} 
- \tau(H_u)
\end{align*}

Summing this quantity over all $u$ yields the exact induced $4$-star count.

\begin{figure*}[!t]
    \centering
    \includegraphics[width=\textwidth]{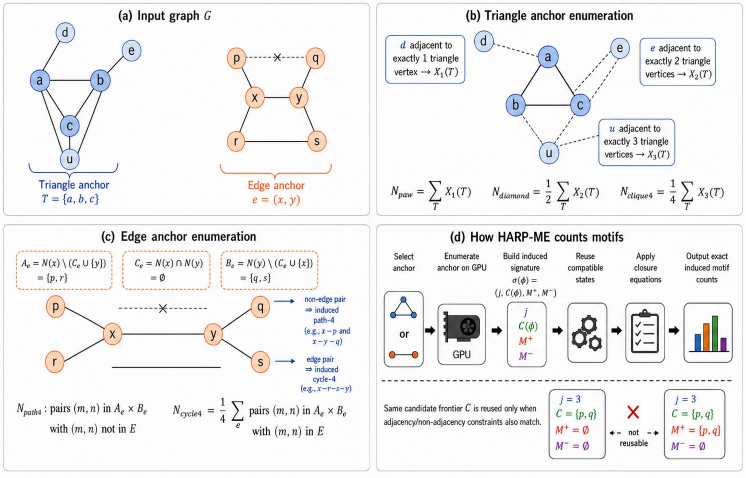}
    \caption{Running example illustrating anchor-based induced 4-node motif enumeration and closure-driven counting.}
    \label{fig:motif}
\end{figure*}

The figure~\ref{fig:motif} illustrates the running example of HARP-ME for exact induced 4-node motif enumeration. First, the system selects smaller structures, such as triangles and edges, as anchors and enumerates them on the GPU. For a triangle anchor, external vertices are classified according to whether they connect to one, two, or all three triangle vertices, allowing paw, diamond, and 4-clique counts to be derived using closure equations. For an edge anchor, candidate vertices on both sides are examined: a non-edge between the candidates forms an induced 4-path, whereas an edge forms an induced 4-cycle. HARP-ME also reuses intermediate states only when both their candidate sets and induced adjacency/non-adjacency constraints match, ensuring correct motif counts while reducing repeated computation.

\section{Experimental Evaluation}\label{Experimental}
All our experiments were run on AQUA cluster. The configuration of each compute node as follows: Intel Xeon Gold 6248 CPU with 40 hardware threads spread over two
sockets, 2.50 GHz clock, and 192 GB memory running RHEL 7.6 OS. All the codes in C++ are compiled with GCC 9.2,
using the optimization flag -O3. We used 
CUDA version 10.1.243 and ran it on the Nvidia Tesla V100-PCIE GPU with 5120 CUDA cores spread uniformly across 80 SMs, clocked at 1.38 GHz with 32 GB global memory and 48 KB shared memory per thread-block.
\subsection{Baselines}
We compare our method against four representative baselines:
\begin{itemize}
    \item \textbf{Pangolin}, a general-purpose GPU graph mining framework.
    \item \textbf{PBE partitioned enumeration}, representing partition-first motif enumeration strategies.
    \item \textbf{Reuse-based GPU enumeration}, capturing prior GPU approaches that exploit intermediate reuse.
    \item \textbf{ORCA-style CPU graphlet baselines}, which are competitive for small motifs on CPUs.
\end{itemize}

\subsection{Datasets}
We have used a set of total six graphs for our experiment. Table~\ref{tab:graph-inputs} represents that set of graphs.
\begin{table}[t]
\centering
\caption{Input graphs.}
\label{tab:graph-inputs}
\begin{tabular}{l|c|r|r}
\hline
\textbf{Graph} &
\textbf{Acronym} &
\multicolumn{1}{c|}{\textbf{$|V|$}} &
\multicolumn{1}{c}{\textbf{$|E|$}} \\
&
&
\multicolumn{1}{c|}{(million)} &
\multicolumn{1}{c}{(million)} \\
\hline

LiveJournal   & LJ & 4.848  & 68.994 \\
soc-Pokec     & PK & 1.633  & 30.623 \\
Web-Google    & WG & 0.876  & 5.105  \\
Web-BerkStan  & WB & 0.685  & 7.601  \\
Bio-CElegans &  BC &  0.000297  &  0.002359  \\
RMAT-24       & RM & 16.777 & 87.600 \\
\hline
\end{tabular}
\end{table}

We evaluate on diverse graph families to capture different sparsity patterns, degree skew, and locality characteristics:
\begin{itemize}
    \item \textbf{SNAP social graphs}, including examples such as Pokec and LiveJournal.
    \item \textbf{Web graphs}, which exhibit strong skew and large intersection frontiers.

  \item \textbf{Biological graphs}, which are relatively sparse and exhibit localized connectivity patterns with lower average degree than social and web graphs.
    \item \textbf{Synthetic RMAT and power-law graphs}, used to stress high-degree vertices and controllable skew.
\end{itemize}

\subsection{Overall Performance}
\label{sec:overall_performance}

Table~\ref{tab:runtime} reports end-to-end runtime across six graphs. HARP-ME is the
fastest method on every evaluated dataset. Relative to Pangolin, the
speedup ranges from $1.36\times$ on Bio-CElegans
($1.9/1.4$) to $2.08\times$ on RMAT-24 ($36.5/17.6$).
The gain is particularly strong on LiveJournal, where runtime decreases
from $42.7$~s to $21.3$~s, a $2.00\times$ speedup, and on Pokec, where
runtime decreases from $18.4$~s to $8.7$~s, a $2.11\times$ speedup.

Compared with partitioned PBE, HARP-ME achieves approximately
$1.50\times$--$1.82\times$ speedup on the larger social, web, and
synthetic graphs. For example, LiveJournal decreases from $37.1$~s to
$21.3$~s ($1.74\times$) and RMAT-24 from $29.4$~s to $17.6$~s
($1.67\times$). These improvements coincide with the lower transfer
fractions reported in Table~3, suggesting that ownership-preserving
anchor statistics reduce partition-induced movement in addition to
explicit enumeration work.

Against Reuse-GPU, HARP-ME improves runtime from $13.8$~s to $8.7$~s
on Pokec ($1.59\times$), from $31.4$~s to $21.3$~s on LiveJournal
($1.47\times$), and from $25.8$~s to $17.6$~s on RMAT-24
($1.47\times$). This comparison is particularly relevant because both
approaches exploit repeated intermediate computation. The remaining
advantage indicates that caching intersections alone does not capture
the full benefit: HARP-ME additionally reduces explicit work through
closure-driven anchor selection and distinguishes reusable states by
induced completion semantics.

The smallest absolute runtimes occur on Bio-CElegans. Here HARP-ME
completes in $1.4$~s, compared with $1.7$~s for Reuse-GPU and $1.9$~s
for Pangolin. The smaller relative gap is consistent with the lower
cache-hit rate and GPU occupancy reported in Table~\ref{tab:micro}, where
Bio-CElegans reaches $64\%$ cache-hit rate and $58\%$ occupancy.
This suggests that small graphs expose less repeated work and less
parallel slack, limiting the benefit of the GPU-specific mechanisms.

The CPU baseline shows the widest gap on large social and synthetic
graphs: HARP-ME reduces LiveJournal runtime from $201.5$~s to $21.3$~s
($9.46\times$), and RMAT-24 from $188.9$~s to $17.6$~s
($10.73\times$). On Bio-CElegans, however, the gap is only
$2.71\times$, again showing that GPU acceleration is most beneficial
when the graph exposes sufficient parallelism and repeated frontier
structure.

Overall, Table~\ref{tab:runtime} supports two conclusions. First, the advantage is not
limited to one graph family: HARP-ME leads on social, web, biological,
and synthetic inputs. Second, the magnitude of improvement increases
on graphs with greater skew and repeated intersection structure, which
is consistent with the mechanism-level measurements examined next.

\begin{table*}[t]
\centering
\caption{Runtime comparison in seconds compared against Proposed method}
\label{tab:runtime}
\begin{tabular}{lrrrrr}
\toprule
Dataset & Pangolin & PBE Part. & Reuse-GPU & ORCA CPU & HARP-ME \\
\midrule
Pokec        & 18.4 & 15.9 & 13.8 &  94.2 &  8.7 \\
LiveJournal  & 42.7 & 37.1 & 31.4 & 201.5 & 21.3 \\
Web-Google   &  9.8 &  8.6 &  7.9 &  34.4 &  5.9 \\
Web-BerkStan & 27.2 & 24.8 & 21.6 & 116.7 & 15.1 \\
Bio-CElegans &  1.9 &  2.1 &  1.7 &   3.8 &  1.4 \\
RMAT-24      & 36.5 & 29.4 & 25.8 & 188.9 & 17.6 \\
\bottomrule
\end{tabular}
\end{table*}

\subsection{Microarchitectural Analysis}
\label{sec:microarchitectural_analysis}
To understand the architectural reasons behind the observed end-to-end performance improvements, we analyze key GPU microarchitectural metrics, including cache-hit rate, memory bandwidth, occupancy, and warp efficiency. This analysis helps identify how HARP-ME's closure-aware anchor selection and induced-signature reuse translate into improved hardware utilization.

Table~\ref{tab:micro} provides mechanism-level evidence for the end-to-end trends.
Cache-hit rate ranges from $64\%$ to $76\%$, with the highest value on
RMAT-24. RMAT-24 also achieves the highest effective bandwidth
($671$~GB/s), occupancy ($84\%$), and warp efficiency ($88\%$).
This combination is consistent with its strong end-to-end performance:
repeated induced states are frequent enough to be reused, while
sufficient anchor parallelism remains available to keep the GPU
occupied.

LiveJournal exhibits a similar pattern, reaching a $74\%$ cache-hit
rate, $645$~GB/s effective bandwidth, $81\%$ occupancy, and $86\%$
warp efficiency. In contrast, Bio-CElegans reaches only $221$~GB/s,
$58\%$ occupancy, and $74\%$ warp efficiency. This difference helps
explain why the relative speedup on the biological graph is smaller
than on the larger social and synthetic graphs.

The embedding and intersection columns should be interpreted as
workload indicators rather than performance metrics in isolation.
LiveJournal processes $884$ million embeddings and $1.21$ billion
intersections, while RMAT-24 processes $703$ million embeddings and
$980$ million intersections. Despite these large workloads, both
graphs sustain high cache-hit rates and GPU utilization. The result
suggests that HARP-ME's advantage is not obtained merely by avoiding
all intersection work; rather, it combines reduced explicit expansion
with reuse of repeated induced completion states.

\begin{table*}[t]
\centering
\caption{Microarchitectural statistics for the proposed method.}
\label{tab:micro}
\begin{tabular}{lrrrrrr}
\toprule
Dataset & Emb. (M) & Inter. (M) & Cache Hit & BW (GB/s) & Occupancy & Warp Eff. \\
\midrule
Pokec        & 312 &  428 & 71\% & 612 & 78\% & 83\% \\
LiveJournal  & 884 & 1210 & 74\% & 645 & 81\% & 86\% \\
Web-Google   &  96 &  133 & 68\% & 534 & 73\% & 79\% \\
Web-BerkStan & 241 &  337 & 70\% & 589 & 76\% & 82\% \\
Bio-CElegans &  12 &   18 & 64\% & 221 & 58\% & 74\% \\
RMAT-24      & 703 &  980 & 76\% & 671 & 84\% & 88\% \\
\bottomrule
\end{tabular}
\end{table*}

\subsection{Partitioning and Transfer Overhead}
\label{sec:partitioning_transfer}

Table~\ref{tab:transfer} shows that HARP-ME substantially reduces the fraction of end-to-end time spent in host--device transfer. On Pokec, transfer overhead decreases from $18.2\%$ under PBE-style partitioning to $6.4\%$, a reduction of $11.8$ percentage points. LiveJournal decreases from $21.5\%$ to $7.1\%$, Web-BerkStan from $16.7\%$ to $5.8\%$, and RMAT-24 from $19.8\%$ to $6.9\%$.

The reduction is consistent across all four partitioned workloads:
HARP-ME's transfer fraction remains below $7.1\%$, whereas PBE ranges
from $16.7\%$ to $21.5\%$. The key distinction is that HARP-ME assigns
ownership to canonical anchor certificates and transfers only the halo information and sufficient statistics required by the selected closure schedule. Replicated halo vertices may therefore support local
completion checks without independently contributing derived counts.

This result matters because partitioning can otherwise erase
kernel-level speedups through repeated data movement. The measurements
indicate that the ownership rule serves both a correctness
role---preventing duplicate derived contributions---and a systems role
by limiting redundant partition-level work.

\begin{table*}[t]
\centering
\caption{Host-device transfer overhead as a percentage of end-to-end runtime.}
\label{tab:transfer}
\begin{tabular}{lrr}
\toprule
Dataset & PBE Transfer Overhead & Proposed Transfer Overhead \\
\midrule
Pokec        & 18.2\% & 6.4\% \\
LiveJournal  & 21.5\% & 7.1\% \\
Web-BerkStan & 16.7\% & 5.8\% \\
RMAT-24      & 19.8\% & 6.9\% \\
\bottomrule
\end{tabular}
\end{table*}

\subsection{Discussion}
The results indicate three main trends. First, reducing the number of set intersections directly lowers total runtime. Second, higher cache hit rate improves effective bandwidth and warp efficiency. Third, compared with partitioned baselines, lower host-device transfer overhead makes the proposed method more robust on large graphs with irregular access patterns.

\section{Related Work}
\label{sec:related_work}

GPU graph pattern mining systems have progressively improved
programmability, scheduling, memory efficiency, and load balancing.
Pangolin~\cite{chen2020pangolin} introduced a high-level
extend-reduce-filter abstraction for efficient graph mining on CPUs and
GPUs. Compilation-oriented systems such as
AutoMine~\cite{mawhirter2019automine},
GraphZero~\cite{mawhirter2021graphzero}, and
GraphPi~\cite{shi2020graphpi} demonstrated that pattern-aware schedule
generation, symmetry breaking, and redundancy elimination can
substantially reduce explicit search. Earlier systems such as
Arabesque~\cite{teixeira2015arabesque},
Fractal~\cite{dias2019fractal}, and
Peregrine~\cite{jamshidi2020peregrine} further established the
importance of high-level pattern abstractions and optimized exploration
of embedding spaces.

Scalability beyond device memory has motivated partitioned and
multi-GPU execution. Partition-based GPU enumeration
~\cite{guo2020partitioned} processes large graphs through
GPU-manageable subgraphs while controlling redundant exploration.
G$^2$Miner~\cite{chen2022g2miner} combines pattern-, input-, and
architecture-aware optimization with generated GPU code and scalable
execution. Reuse-oriented enumeration~\cite{guo2023reuse} further shows
that repeated set intersections are a major source of redundant work and
can be recycled across related search states. More recently,
DuMato~\cite{dumato2024} reinforced the effectiveness of warp-centric
traversal and dynamic workload balancing for irregular GPU subgraph
enumeration.

HARP-ME is complementary to these systems but differs in two respects.
First, it optimizes which anchor structures should be explicitly
enumerated rather than only how a fixed pattern should be traversed.
Second, induced-state reuse requires agreement in both adjacency and
non-adjacency constraints; identical candidate frontiers alone are not
sufficient. HARP-ME therefore combines closure-aware anchor selection
with induced-signature reuse and ownership-preserving partitioned
execution.

A related direction avoids direct execution of every requested pattern.
Subgraph Morphing~\cite{jamshidi2023subgraphmorphing} transforms costly
pattern computations into alternative structures and reconstructs the
desired results. Combinatorial graphlet methods similarly exploit
relations among small patterns. ORCA~\cite{hocevar2014orca} derives
graphlet orbit counts through systems of equations,
ESCAPE~\cite{pinar2017escape} uses combinatorial identities and local
statistics for efficient five-vertex subgraph counting, and
PGD~\cite{ahmed2015pgd} exploits structural decomposition for scalable
graphlet counting. These works motivate reducing direct enumeration
through algebraic structure.

HARP-ME brings these directions together in a GPU framework for exact induced motifs. Unlike traversal-only systems, its compilation objective jointly considers explicit enumeration cost, repeated induced-state reuse, closure yield, and partition-induced halo amplification.

\section{Conclusion and Future Work}\label{conclusion}

We presented HARP-ME, a GPU framework for exact connected induced
4-node motif enumeration that combines closure-aware anchor selection,
induced-signature reuse, and ownership-preserving partitioned execution.
By explicitly enumerating a selected anchor basis and deriving residual
counts through exact closure relations, HARP-ME reduces redundant
expansion and repeated induced checks.

Across diverse graphs, HARP-ME achieved up to 2.11$\times$
speedup over Pangolin, 1.83$\times$ over partitioned PBE, and
10.73$\times$ over the CPU baseline, while reducing host-device
transfer overhead. These
results show that closure yield can effectively complement
traversal- and reuse-oriented GPU optimizations.

Future work will extend HARP-ME to larger motif families, including
5-node and higher-order induced motifs, and investigate automatic
closure synthesis, adaptive anchor selection, and scalable multi-GPU
execution.

\bibliographystyle{IEEEtran}
\bibliography{sample-base}

\end{document}